\documentstyle[11pt,epsfig]{article}

\begin{document}

\begin{titlepage}

\bigskip

\bigskip

\begin{center}{\Large The search for charge sleptons and
flavour lepton number violation at LHC (CMS)}
\end{center}

\bigskip

\begin{center}
{\large
    S.I.~Bityukov (IHEP, Protvino RU-142284, Russia),\\
    N.V.~Krasnikov (INR, Moscow 117312, Russia)
}
\end {center}

\vspace{2cm}
  
\begin{center}

{\large Presented at the International Seminar QUARKS-98\\
May 18-24, 1998, Suzdal, Russia}

\end {center}

\vspace{2cm}
  
\begin{abstract}

We study a possibility to detect charged sleptons and flavour lepton number
violation at LHC (CMS). We investigate the production and
decays of right- and left-handed sleptons separately. We have found that for
luminosity
$L~=~10^5pb^{-1}$ it would be possible to detect right-handed sleptons
with a mass up to 325~GeV and left-handed ones with a mass up to 350~GeV.
We also investigate a possibility to look for flavour lepton number
violation in slepton decays due to the mixing of different sleptons
generations. We find that for maximal $(\tilde \mu_{R}-\tilde e_{R})$
mixing sleptons detection is possible up to masses of about 250~GeV.

\end{abstract}

\vspace{4cm}

\end{titlepage}

\section{Introduction}

This talk is based on our early paper (S.I.~Bityukov and N.V.~Krasnikov,
Preprint IFVE 97-67, also hep-ph/9712358).

One of the LHC goals is the discovery of the supersymmetry. 
In particular, it is very important to investigate a possibility
to discover nonstrongly interacting superparticles (sleptons,
higgsino, gaugino). In ref.\cite{1,2,3} the LHC slepton discovery potential
has been investigated within the minimal SUGRA-MSSM framework where all 
sparticle masses are determined mainly by two parameters $m_0$ (common
squark and slepton mass at GUT scale) and $m_{1 \over 2}$ (common
gaugino mass at GUT scale). 
The signature used for the search for sleptons at LHC is 
$two~same-flavour~opposite-sign~leptons + E^{miss}_T + no~jets$~\cite{1,2,3}. 
The conclusion of these studies is that LHC is able 
to detect sleptons with the masses up to (300-400)~GeV.

In this paper we investigate the discovery potential
for sleptons and for flavour lepton number violation in slepton decays at CMS.
This study complements the previous ones~\cite{1,2,3} as we do not 
use the minimal SUGRA-MSSM framework. Instead, we investigate separately 
the production and decays of right-handed and left-handed sleptons. Despite
the simplicity of the  SUGRA-MSSM framework it is a very particular
model. The mass formulae for sparticles in  SUGRA-MSSM model are derived
under the assumption that at GUT scale ($M_{GUT} \approx 2 \cdot 10^{16}$~GeV) 
soft supersymmetry breaking parameters are universal. However, in general,
we can expect that real sparticle masses can differ in a drastic way 
from sparticle masses pattern of SUGRA-MSSM model due to many reasons:

\begin{enumerate}
\item
in superstring inspired models soft scalar supersymmetry breaking terms
are not universal at Planck scale in general~\cite{4},
\item
in supersymmetric SU(5) model an account of the evolution of soft 
supersymmetry breaking terms between Planck and GUT scale~\cite{5,6} 
is very essential,
\item 
in models with additional relatively light vector like supermultiplets
the mass formulae for superparticles can drastically differ~\cite{7}
from the standard ones~\cite{8}.
\end{enumerate}

\noindent
Therefore, it is more appropriate to investigate the LHC SUSY sensitivity
in a model-independent way. The cross section for the production of the
right(left)-handed sleptons depends mainly on the mass of right(left)-handed
sleptons and the decay properties of the sleptons are determined mainly 
by the mass of the lightest superparticle (LSP).

Thus, the discovery potential of the right(left)-handed sleptons depends
mainly on 2 parameters -- the slepton mass and the LSP mass. Here
we investigate the possibility to look for right(left)-handed
sleptons for the case of arbitrary masses of right(left)-handed sleptons
and LSP. We also investigate the possibility to look for flavour lepton 
number violation in slepton decays at LHC. 

For the LHC slepton search we use the signature:

$two~same~flavour~with~the~opposite~charged~leptons + E^{miss}_T + no~jets$. 
There are two types of backgrounds for this signature: standard model 
background and SUSY 
strong $(\tilde g \tilde g,~\tilde g \tilde q,~\tilde q \tilde q)$ and 
weak
$(\tilde \chi^{\pm}_1 \tilde \chi^0_2,~\tilde \chi^{\pm}_1 \tilde \chi^{\mp}_1)$ backgrounds.
As a rule the SUSY strong  and weak backgrounds are not very large~\cite{3} 
and they only increase the LHC discovery potential of new SUSY physics
(however, in general, it could be nontrivial to separate a slepton signal
from SUSY backgrounds).

Our simulations are made at the particle level with parametrized
detector responses based on a detailed detector simulation. The CMS detector
simulation program CMSJET~3.2~\cite{11} is used.
All SUSY processes with full particle spectrum, couplings,
production cross section and decays are generated with ISAJET~7.13,
ISASUSY~\cite{9}. The Standard Model backgrounds are generated 
with PYTHIA~5.7~\cite{10}. We have used the same cuts and estimates for the
Standard Model backgrounds obtained in ref.~\cite{3}. 

In Section~2 we give short review of sleptons in MSSM.
In Section~3 we describe slepton production and decay mechanisms.
Section~4 is devoted to the discussion of the Standard Model backgrounds.
In Sections~5, 6 and 7 we discuss the case of right-handed,
left-handed and left- plus right-handed sleptons, correspondingly.
Section~8 is devoted to the discussion of the search for flavour lepton
violation in slepton decays. Section 9 contains concluding remarks.

\section{Sleptons in MSSM framework}

In MSSM framework all gaugino masses coincide at GUT scale
$M_{GUT} \approx 2 \cdot 10^{16}GeV$. As a result of the evolution of the 
effective masses between GUT and electroweak scales the relation between the
chargino and neutralino masses is~\cite{8}

\begin{center}
$m(\tilde \chi^0_2) \approx m(\tilde \chi^{\pm}_1) \approx 2~m(\tilde \chi^0_1) 
\approx m_{1 \over 2},$
\end{center}
where $m_{1 \over 2}$ is common gaugino mass at GUT scale.

In MSSM slepton masses are determined by formulae~\cite{8}:

\begin{equation}
m^2_{\tilde l_R} = m^2_0 + 0.15~m^2_{1 \over 2} - 
sin^2\theta_WM^2_Zcos2\beta
\end{equation}

\begin{equation}
m^2_{\tilde l_L} = m^2_0 + 0.52~m^2_{1 \over 2} - 
{1 \over 2}(1 - 2~sin^2\theta_W)M^2_Zcos2\beta
\end{equation}

\begin{equation}
m^2_{\tilde \nu} = m^2_0 + 0.52~m^2_{1 \over 2} - 
{1 \over 2}M^2_Zcos2\beta,
\end{equation}

\noindent
where $m_0$ is the common scalar soft breaking mass at GUT scale.
As it follows from formulae (1-3) right-handed sleptons are lighter
than left-handed ones, i.e. $m_{\tilde l_R} < m_{\tilde l_L}$.

In general the decays of sleptons can be rather complicated.
For the case when right-handed sleptons are the lightest among non LSP
sparticles they decay dominantly to LSP
\begin{center}
$\tilde l^-_R \longrightarrow l^- + \tilde \chi^0_1$.
\end{center}

\noindent
The left-handed sleptons decay (if kinematically accessible) 
to charginos and neutralinos\\

\hspace{4cm} $\tilde l^{\pm}_L \longrightarrow l^{\pm} + \tilde \chi^0_{1,2}$

\hspace{4cm} $\tilde l^{\pm}_L \longrightarrow \nu_L + \tilde \chi^{\pm}_1$

\hspace{4cm} $\tilde \nu_L \longrightarrow \nu_L + \tilde \chi^0_{1,2}$

\hspace{4cm} $\tilde \nu_L \longrightarrow l^{\pm} + \tilde \chi^{\mp}_1.$\\

\vspace*{0.2cm}

\noindent
Slepton pairs ($\tilde l_L~\tilde l_L,~\tilde l_R~\tilde l_R,~\tilde 
\nu_L~\tilde \nu_L,~\tilde \nu_L~\tilde l_L$) 
can be produced either through a Drell - Yan mechanism or,
if kinematically allowed through the decays of other supersymmetric
particles ($\tilde \chi^0_2,~\tilde \chi^{\pm}_1,~...$). 
For the case when $\tilde \chi^0_2,~\tilde \chi^{\pm}_1$ are heavier than sleptons,
an indirect slepton production is open:

\hspace{4cm} $\tilde \chi^0_2 \longrightarrow \tilde l^{\pm}_{L,R}l^{\mp}$

\hspace{4cm} $\tilde \chi^0_2 \longrightarrow \tilde \nu_L \bar \nu_L$

\hspace{4cm} $\tilde \chi^{\pm}_1 \longrightarrow \tilde \nu_L l^{\pm}$

\hspace{4cm} $\tilde \chi^{\pm}_1 \longrightarrow \tilde l^{\pm}_L \nu_L.$\\

\vspace*{0.2cm}

Right-handed and left-handed charged
sleptons decay to LSP and leptons 

\hspace{4cm}$\tilde l^-_R \longrightarrow l^- + \tilde \chi^0_1$,

\hspace{4cm}$\tilde l^-_L \longrightarrow l^- + \tilde \chi^0_1$,\\
leading to the signature:
$two~same-flavour~opposite-sign~leptons + E^{miss}_T + no~jets$. 
This signature can be realized also as a result of the gaugino decays:

\vspace*{0.2cm}
\hspace{4cm} $\tilde \chi^0_2 \longrightarrow \tilde \chi^0_1 + l^+l^-$

\hspace{4cm} $\tilde \chi^0_2 \longrightarrow \tilde \chi^0_1 + \nu \bar \nu$

\hspace{4cm} $\tilde \chi^0_2 \longrightarrow \tilde \chi^0_1 + Z^0$

\hspace{4cm} $\tilde \chi^{\pm}_1 \longrightarrow \tilde \chi^0_1 + l^{\pm} + \nu$

\hspace{4cm} $\tilde \chi^{\pm}_1 \longrightarrow \tilde \chi^0_1 + W^{\pm}.$

\vspace*{0.2cm}

\section{Sleptons production and decays}

In this paper we study only
direct production of charged sleptons via the Drell - Yan mechanism with
their subsequent decays into leptons and LSP.
We study the case when right-handed and left-handed charged
sleptons decay to LSP and leptons 

\begin{center}
$\tilde l^-_R \longrightarrow l^- + \tilde \chi^0_1$,\\

$\tilde l^-_L \longrightarrow l^- + \tilde \chi^0_1$.\\
\end{center}
The simulation when the slepton decay dominantly to LSP and leptons 
correspond to the case when chargino and second neutralino are 
heavier than sleptons. In our study we neglect indirect slepton 
production. Inclusion of indirect slepton production or other SUSY backgrounds
only increases the excess of the supersymmetric signal over
SM background and enhances the value of significance.

\section{Standard Model backgrounds}

The expected main Standard Model background process should be $t \bar t$
production, with both W's decaying to leptons, or one of the leptons 
from W decay and the other from the $b-$decay of the same $t-$quark;
the other SM backgrounds come from WW, WZ, $b \bar b$ and 
$\tau \tau$-pair production, with decays to electrons and muons.
Standard model backgrounds for different kinematical cuts have been 
calculated~\cite{3} with the PYTHIA~5.7 code. In this paper
we use the results from ref.~\cite{3}. We also considered possible
backgrounds from 

\begin{center}
$p p \rightarrow Z~Z \rightarrow \nu\nu~~\bar \tau \tau$,\\

$p p \rightarrow Z~jet \rightarrow \bar \tau \tau~~jet$ 
\end{center}

processes and have found that they are small. These backgrounds give less 
than two events for $L~=~10^5pb^{-1}$.

The set of kinematical variables which are useful to extract the slepton
signals and typical selection cuts are~\cite{3}:
\begin{itemize}
\item[i)] for leptons ($l=e,\mu$)~:

\begin{itemize}
\item
$p_T-$cut on leptons and lepton isolation (Isol), which is here defined 
as the calorimetric energy flow around the lepton in a cone
$\Delta R~<~0.5$ divided by the lepton energy;
\item 
mass of the same-flavour opposite-sign leptons $M_{l^+l^-} \neq M_Z$ 
to suppress $WZ$ and potential $ZZ$ backgrounds by rejecting events in
a $M_Z \pm \delta M_Z$ band;
\item
$\Delta \Phi(l^+l^-)-$relative azimuthal angle between two same-flavour
opposite-sign leptons in plane transverse to the beam;
\end{itemize}

\item[ii)] for $E^{miss}_T~:$

\begin{itemize}
\item
$E^{miss}_T-$cut,
\item
$\Delta \Phi(E^{miss}_T,ll)-$relative azimuthal angle between 
$E^{miss}_T$ and the resulting dilepton momentum in the transverse plane;
\end{itemize}

\item[iii)] for jets~:

\begin{itemize}
\item
"jet veto"-cut~: $N_{jet}~=~0$ for some $E^{jet}_T$ threshold,
in some rapidity interval, typically $|\eta_{jet}|~<~4.5$~(we use
standard $UA1$ jet definition with $R_{cone}~=~0.5.$

\end{itemize}

\end{itemize}

Namely, we adopt from the ref.~\cite{3} the sets of cuts
which in our notation looks as follows:

\begin{table}[h]
\small
 \caption{The cut values of kinematical variables for sets (1-7) of cuts
and the distribution of SM backgrounds events for each set of cuts\cite{3}.}
    \label{tab:Tab.1}
\begin{center}
\begin{tabular}{|l|l|l|l|l|l|l|l|}
\hline
Cut~~$\backslash$~~Set~$\#$ & 1 & 2 & 3 & 4 & 5 & 6 & 7 \\
\hline
$p^l_T~>$  & 20~GeV & 20~GeV &50 GeV&50 GeV& 60 GeV& 60 GeV& 60 GeV\\
$Isol~<$   &    0.1 &  0.1   & 0.1 &0.1&0.1&0.1&0.03\\
$|\eta_l|~<$             &    2.5 &  2.5   & 2.5 &2.5&2.5&2.5&2.5\\
\hline
$E^{miss}_T~>$  & 50~GeV & 50~GeV & 100 GeV&120 GeV&150 GeV&150 GeV&150 GeV\\
\hline
$\Delta \Phi(E^{miss}_T,ll)>$ &
$160^o$& $160^o$  &$150^o$&$150^o$&$150^o$&$150^o$&$150^o$\\
\hline
$N_{jet}~=$              &      0 &  0     & 0& 0& 0&0&0\\
$E^{jet}_T~>$      & 30 GeV & 30 GeV & 30 GeV&30 GeV&45 GeV&45 GeV&45 GeV\\
$|\eta_{jet}|~<$      &    4.5 &    4.5 & 4.5&4.5&4.5&4.5&4.5\\
\hline
$M_Z-cut$& yes & yes  & yes & yes & yes & yes & yes \\
\hline
$\Delta \Phi(l^+l^-)$  &  
$>130^o$ & no &$<130^o$ &$<130^o$&$<130^o$&$<140^o$&$<130^o$\\
\hline
$WW~\cite{3} ; ~~\sigma_{WW}=70pb$ & 454 & 1212 & 97  & 69  & 38 & 46 & 32 \\
$Wt\bar b~\cite{3} ; ~~\sigma_{Wt\bar b}=160pb$   & 163 & 577  & 33  & 13  &  1 & 1  &  1 \\
$t\bar t~\cite{3} ; ~~\sigma_{t\bar t}=660pb$    & 345 & 574  & 21  &  6  &  0 & 0  &  0 \\
$WZ~\cite{3} ; ~~\sigma_{WZ}=26pb$       & 15  & 43   & 21  & 17  &  6 & 6  &  5 \\
$\tau \tau~\cite{3} ; ~~\sigma_{\tau \tau}=7.5pb$  & 15  & 15   &  0  &  0  &  0 & 0  &  0 \\
\hline
$N^{SM}_B$~\cite{3}    & 992 & 2421 & 172 & 105 & 45 & 53 & 38 \\
\hline
\end {tabular}
\end{center} 
\end{table}

\normalsize

\noindent
Here $N^{SM}_B$ is the number of the Standard Model background events
for $L~=~10^4pb^{-1}$ (cuts 1-2) and for $L~=~10^5pb^{-1}$ (cuts 3-7).
$M_Z-cut$ is condition that $M_{l^+l^-}~<~86~GeV$ or $96~GeV~<~M_{l^+l^-}$.
$CTEQ2L$ parton distributions have been used. Note that the sets of cuts (1-2)
are effective for the search for relatively light sleptons 
$(m_{\tilde l} \leq 150~GeV)$ whereas the sets (3-7) are appropriate for the
heavy sleptons with $(m_{\tilde l} > 150~GeV)$.

As has been mentioned above, we neglect indirect
slepton production and also we neglect SUSY backgrounds which are mainly due to
$\tilde q \tilde q,~\tilde g \tilde q,~\tilde g \tilde g$ production and with
subsequent cascade decays with jets outside the acceptance or below the
threshold.
As has been demonstrated in ref.~\cite{3} 
SUSY background is, as a rule, much less than the SM background and we shall
neglect it.

\section{Right-handed sleptons}

In this section we study the possibility to search for right-handed sleptons
at CMS. Namely, we consider the signature {\it dilepton +
$E^{miss}_T$~+~no~jets}. We do not consider the left-handed slepton
contribution to this signature, i.e. we consider the situation when left-handed
sleptons are much heavier than the right-handed ones 
and it is possible to neglect
them. In this case right-handed sleptons decay dominantly to an LSP

\begin{center}
$\tilde l^-_R \rightarrow l^- + \tilde \chi^0_1$.
\end{center}

\noindent
The cross section of the right-handed slepton production is determined
mainly by the mass of the right-handed slepton. The dependence of the
right-handed slepton cross section production for the case of 3-flavour
degenerate right-handed charged sleptons as a function of the slepton mass 
is presented in Table~2 and in Fig.1. 
Comparison of shapes of $E_T^{miss}$ distribution and $p_T^l$ of leptons 
distribution in case of charged
right-handed sleptons ($m_{\tilde l_R}$~=~200~GeV,~~$m_{\tilde \chi^0_1}~=53~GeV$) 
with background is shown in Fig.2.

\begin{table}[h]
\small
 \caption{The cross section $\sigma(p p \rightarrow \tilde l^+_R \tilde l^-_R
\, + \, ...)$ in pb for different values of charged right-handed slepton 
masses for 3 slepton generations at LHC.
Right-handed sleptons are assumed to be degenerate in mass.}
    \label{tab:Tab.2}
    \begin{center}
\begin{tabular}{|l|l|l|l|l|l|l|l|}
\hline
$M(GeV)$ &    90 &   100  &   125  &   150  &   175  &   200  &   225 \\ 
\hline
$\sigma$ &  0.41& 0.27  & 0.13  &  0.068 & 0.039 &  0.024 &  0.016 \\
\hline
\hline
$M(GeV)$ &  250  &   275  &   300  &   325  &   350  &   375  &   400 \\ 
\hline
$\sigma$ &0.011 &0.0079 & 0.0055& 0.0041& 0.0032 & 0.0025& 0.0020 \\
\hline
\end{tabular}
    \end{center}
\end{table}

\noindent
The sum of distributions of the SM backgrounds and signal events
after set 7 for $E_T^{miss}$~(a) and $p_T^l$ of leptons~(b) in case of
$m_{\tilde l_R}$~=~200~GeV,~~$m_{\tilde \chi^0_1}~=53~GeV$ is shown in Fig.3.
The number of signal events
passing through the cuts 1-7 essentially depends on the LSP mass
$m_{\tilde \chi^0_1}$. 
The results of our calculations for different values of
the slepton and the LSP masses are presented in Tables~$4-15$. In Tables
the significance $S$ is defined as \cite{3}

\begin{equation}
S = \frac{N_S}{\sqrt{N_S~+~N^{SM}_B}},
\end{equation}

\noindent
where $N^{SM}_B$ have been calculated in ref.~\cite{3}.
The meaning of the definition (4) is the following. Suppose the 
background and signal cross sections are $\sigma^{SM}_{B}$ and $\sigma_{S}$ 
correspondingly. For a given luminosity $L$ the average number of events 
expected in the experiment is
$N_{ev} = N_S + N^{SM}_B = (\sigma_{S} + \sigma^{SM}_{B})L$. The fluctuation 
of this number is $\sqrt{N_S + N^{SM}_{B}}$. So the definition (4) takes into 
account statistical uncertainty in the determination of the total number of 
events. Standard definition of the significance $S = \frac{N_S}{\sqrt{N_B}}$ 
corresponds to the situation with real experiment when the total 
number of events $N_{ev} = N^{SM}_B + N_{S}$ is given number and we 
have to compare $N_{ev}$ with the expected number of background events 
$N^{SM}_B$. 

Besides statistical error we have systematical error related with nonexact 
knowledge of the background cross section. We assume that the accuracy 
in the determination of background cross section is between 
10 and 50 percents. Therefore the ratio $\frac{N_S}{N_{B}^{SM}}$ 
at least has  to be bigger than 0.1.   

As it follows from Tables~4-15 for $L~=~10^5pb^{-1}$,
it is possible to detect the right-handed sleptons at 5~$\sigma$
significance level with a mass up to 325~GeV (Table 14, 
$m_{\tilde \chi^0_1}=85~GeV$, cut 6). For the right-handed sleptons
with a mass $90~GeV~\leq~m_{\tilde l_R}~\leq~300~GeV$ it is possible
to discover sleptons for not very large values of the LSP mass. Typically,
$m_{\tilde \chi^0_1}~\leq~(0.4~-~0.6)m_{\tilde l_R}$ should be chosen.

It should be noted that the SM background coming from
$W~W,~W~t~\bar b,~W~Z,~\bar t~t,~\bar \tau~\tau$ 
production with subsequent  leptonic decays 
predicts the equal number of $\mu^+~\mu^-,~e^+~e^-,~\mu^+~e^-$ and 
$e^+~\mu^-$ events up to statistical fluctuation whereas the signal 
contains an equal number of $\mu^+~\mu^-$ and $e^+~e^-$ pairs coming from

\begin{center}
$p~p \rightarrow \tilde \mu^+_R~\tilde \mu^-_R~+~\dots~
\rightarrow \mu^+~\mu^-~+~2~LSP~+~\dots$ 
\end{center}
and

\begin{center}
$p~p \rightarrow \tilde e^+_R~\tilde e^-_R~+~\dots~\rightarrow e^+~e^-~+~
2~LSP~+~\dots~.$
\end{center}

It should be noted that the criterion based on the measurement of 
$\Delta N = N_{ev}(e^+e^- + \mu^{+}\mu^{-}) - N_{ev}(e^{+}\mu^{-} + 
e^{-}\mu^{+}) = N_{S}$ is free from the systematical uncertainty 
related with the nonexact knowledge of the background cross section. 
The $1 \sigma$ statistical fluctuation of $\Delta N$ is 
$\sqrt{1 N^{SM}_{B}}$.

\noindent
We have found also that the reaction 

\begin{center}
$p~p \rightarrow \tilde \tau^+_R~\tilde \tau^-_R~+~\dots~
\rightarrow \mu^+~\mu^-,~e^+~e^-,~e^+~\mu^-,~\mu^+~e^-~+~\dots$
\end{center}

\noindent
virtually does not contribute to the number of signal events and can be 
neglected for the considered signature. Then we study the production and 
decays of the first two generations of sleptons. Therefore, we have 
qualitatuve consequence of the existence of slepton signal -- the excess of
$\mu^+ \mu^-$ and $e^+ e^-$ events over $\mu^+ e^-$ and $e^+ \mu^-$ events.
It allows to apply sequential analysis in observation of difference between
two processes (one of them gives $\mu^+ \mu^-$ and $e^+ e^-$ events, the other
gives $\mu^+ e^-$ and $e^+ \mu^-$ events).    
  This is an additional criterion for new physics discovery for the case 
of standard slepton production, which also allows to eliminate systematics 
related with the nonexact knowledge about background.

\begin{figure}[htpb]
  \begin{center}
    \resizebox{7cm}{!}{\includegraphics{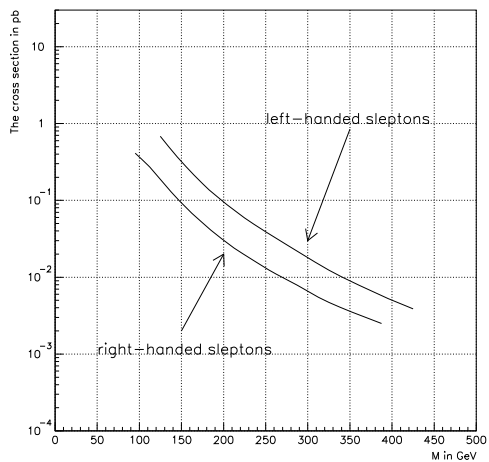}} 
\caption{The cross section $\sigma(p p \rightarrow \tilde l^+_R \tilde l^-_R
\, + \, ...)$ in pb for different values of charged right-handed slepton 
masses at LHC
(right-handed sleptons are assumed to be degenerate in mass) and
the cross section $\sigma(p p \rightarrow \tilde l^+_L \tilde l^-_L
\, + \, ...)$ in pb for different values of charged left-handed slepton masses 
for at LHC (left-handed sleptons masses are assumed to be 
degenerate in flavour).}
    \label{fig:1} 
  \end{center}
\end{figure}

\begin{figure}[htpb]
  \begin{center}
    \resizebox{7cm}{!}{\includegraphics{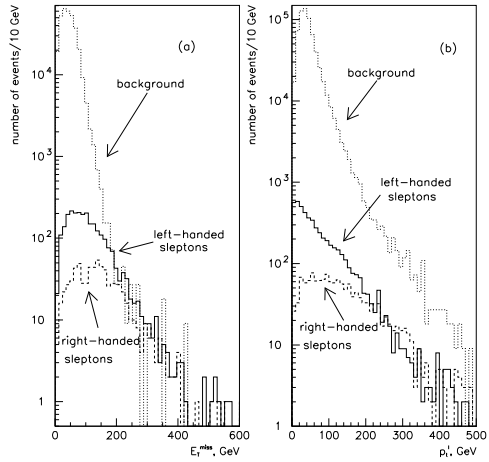}} 
\caption{$E_T^{miss}$ distributions~(a) and $p_T^l$~of~leptons~(b) 
distributions for sleptons and backgrounds before applying 
of sets (1-7) of cuts. 
Here $m_{\tilde l_R}~=~200~GeV,~~m_{\tilde \chi^0_1}~=53~GeV$ in case 
of right-handed sleptons and $m_{\tilde l_L}~=~200~GeV,~~m_{\tilde \chi^0_1}~=53~GeV$ 
for left-handed sleptons.}
    \label{fig:2} 
  \end{center}
\end{figure}
        
\begin{figure}[htpb]
  \begin{center}
    \resizebox{12cm}{!}{\includegraphics{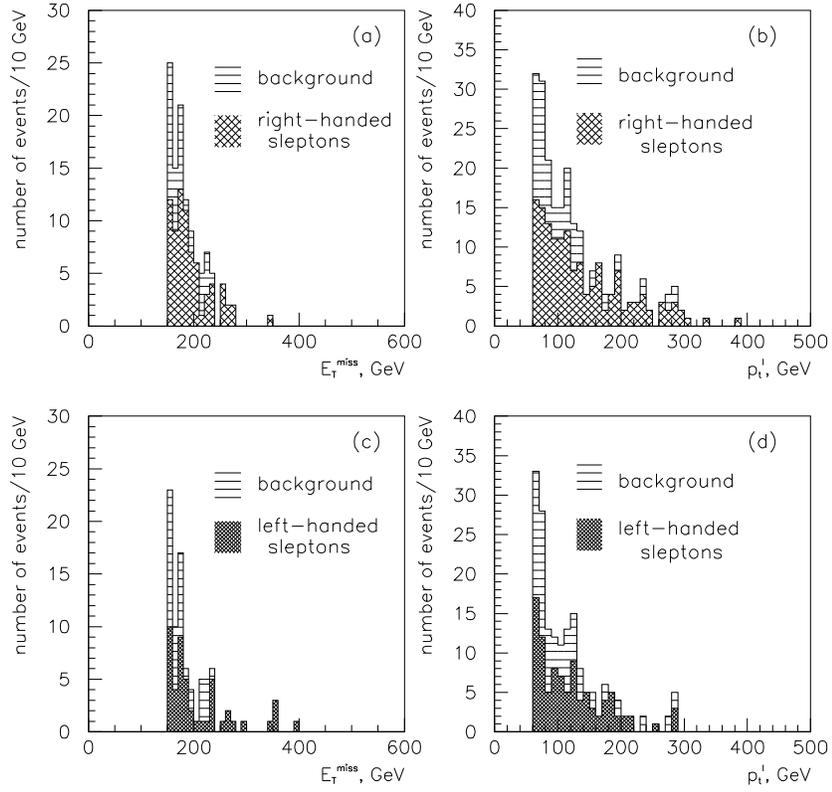}} 
\caption{The sum of distributions SM backgrounds and signal events
after cut 7 for $E_T^{miss}$~(a,c) and $p_T^l$ of leptons~(b,d). 
Here $m_{\tilde l_R}~=~200~GeV,~~m_{\tilde \chi^0_1}~=53~GeV$ in case 
of right-handed sleptons and $m_{\tilde l_L}~=~200~GeV,~~m_{\tilde \chi^0_1}~=53~GeV$ 
for left-handed sleptons.}
    \label{fig:3} 
  \end{center}
\end{figure}

\section{Left-handed sleptons}

In this section we study the case when the right-handed sleptons are much
heavier than the left-handed ones (of course, this case looks
pathological since in MSUGRA approach the left-handed sleptons are
heavier than the right-handed ones, however, in a general case we can not
exclude such a possibility) and we can neglect them. 
The dependence of the 
left-handed slepton cross section production 
on slepton mass is presented in Table~3 
and in Fig.1. Our results
are presented in Tables~16-21. The notation is similar to the case
of the right-handed sleptons. Correspondingly, comparison of shapes of
$E_T^{miss}$ and $p_T^l$ distributions 
($m_{\tilde l_L}$~=~200~GeV,~~$m_{\tilde \chi^0_1}~=53~GeV$) 
with background is shown in Fig.2 and 
the sum of distributions SM backgrounds and signal events
after cut 7 for $E_T^{miss}$~(c) and $p_T^l$ of leptons~(d) in case of
$m_{\tilde l_L}$~=~200~GeV,~~$m_{\tilde \chi^0_1}~=53~GeV$ is shown in Fig.3.
As follows from Tables~16-21, it is
possible to detect left-handed sleptons with a mass up to  350~GeV
(Table 21, $m_{\tilde \chi^0_1}=196~GeV$, cut 6).  
The discovery potential of the left-handed 
sleptons depends, as in the case of the right-handed sleptons, on the mass
of LSP. The LSP masses $m_{\tilde \chi^0_1}~=~(0.4-0.6)m_{\tilde l_L}$ give the
maximal number of events which passed cuts unlike the case of the right-handed 
sleptons when small LSP masses are the most preferable as far as detection 
of LHC sleptons is concerned. 

\begin{table}[h]
\small
 \caption{The cross section $\sigma(p p \rightarrow \tilde l^+_L \tilde l^-_L
\, + \, ...)$ in pb for different values of charged left-handed slepton 
masses at LHC.
Left-handed sleptons masses are assumed to be degenerate in flavour.}
    \label{tab:Tab.3} 
\begin{center}
\begin{tabular}{|l|l|l|l|l|l|l|l|l|l|}
\hline
$M(GeV)$ &  100 &  150 &  200  &   250 &  300  &  350  &  400  & 450\\
\hline
$\sigma$ &0.6830&0.1651&0.05921&0.02644&0.01238&0.00665&0.00389&0.00229 \\
\hline 
\end{tabular}
\end{center}
\end{table}

\section{Right-handed plus left-handed sleptons}

As has been mentioned in the Introduction, in general, we can expect
that MSUGRA model, in the best case, gives qualitative description
of the sparticle spectrum. So, in general, masses of the LSP, the right-handed 
and the left-handed sleptons are arbitrary. In many models the 
left-handed sleptons are heavier than the right-handed ones. 
With a good accuracy the right-handed and the left-handed sleptons give 
an additive contribution to the signal event, i.e.  

\begin{center}
$N(signal)~=~N_{Left}(signal)~+~N_{Right}(signal).$
\end{center}

To obtain some flavour we have studied as an example 
the direct charged sleptons production
for the case when $m_{\tilde l_L}~=~m_{\tilde l_R}~+~50$~GeV.
The results of our investigation are presented in Tables~22-25.
As follows from Tables~22-25, it is possible to discover
sleptons with a mass of the right-handed slepton up to 350~GeV (see Table~25).
The inclusion of the left-handed sleptons increases the slepton
discovery potential; moreover, it is possible to discover sleptons in 
a wide range of LSP masses. Again the criterion based on the difference 
between
$(e^+e^-+\mu^+\mu^-)$ and $(e^+\mu^-+\mu^+e^-)$ events gives some additional 
information about the existence of new physics related to slepton production.

\section{The search for flavour lepton number violation in slepton decays}

As has been mentioned above, in MSUGRA the scalar soft supersymmetry 
breaking terms are postulated to be universal at GUT scale. For such
"standard" supersymmetry breaking terms the lepton flavour number is
conserved in supersymmetric extension of the Weinberg - Salam model.
However, in general, squark and slepton supersymmetry breaking mass terms 
are not diagonal due to many reasons~\cite{12} (for instance, in SU(5)
SUSY GUT model an account of the evolution between Planck and GUT scales
leads to the appearance of nondiagonal soft supersymmetry breaking terms)
 and flavour lepton number
is explicitly broken due to nondiagonal structure of slepton soft 
supersymmetry breaking mass terms. As a consequence such models predict
flavour lepton number violation in $\mu-$ and $\tau-$ decays~\cite{12}.
In ref.~\cite{13,14,15} it has been proposed to look for flavour lepton
number violation in slepton decays at LEP2 and NLC. In ref.~\cite{16}
the possibility to look for flavour lepton number violation in slepton decays 
at LHC has been
studied with 2 points of ref.~\cite{3}. It has been shown that CMS/ATLAS 
detectors at LHC will be able to discover flavour lepton number violation
in slepton decays for the case of maximal mixing up to 
$m_{\tilde l}~\approx~200~GeV$.

In this paper we investigate this signature in a more careful way.
We consider the case of mixing between the right-handed
selectron and the right-handed smuon. The mass term for the right-handed
$\tilde e_R'$ and $\tilde \mu_R'$ sleptons has the form

\begin{equation}
-\Delta {\cal L}~=~m^2_1 \tilde e^{+\prime}_R \tilde e^{\prime}_R~+~
m^2_2 \tilde \mu^{+\prime}_R \tilde \mu^{\prime}_R~+~
m^2_{12}(\tilde e^{+\prime}_R \tilde \mu^{\prime}_R~+
~\tilde \mu^{+\prime}_R \tilde e^{\prime}_R).
\end{equation}

\noindent
In formula (5) the last term explicitly violates lepton flavour number.
After the diagonalization of the mass term~(5), we find that the eigenstates of 
the mass term (5) are

\begin{equation}
\tilde e_R~=~\tilde e_R' cos(\phi)~+~\tilde \mu_R' sin(\phi),
\end{equation}

\begin{equation}
\tilde \mu_R~=~\tilde \mu_R' cos(\phi)~-~\tilde e_R' sin(\phi),
\end{equation}

\noindent
with the masses

\begin{equation}
M^2_{12}~=~\frac{1}{2} [(m^2_1 + m^2_2) \pm 
(~(m^2_1 - m^2_2)^2 + 4(m^2_{12})^2~)^{\frac{1}{2}}],
\end{equation}

\noindent
which coincide virtually for small values of $m^2_1-m^2_2$ and $m^2_{12}.$
Here the mixing angle is determined by the formulae 

\begin{equation}
tan(2 \phi)~=~\frac{2 m^2_{12}}{m^2_1-m^2_2}.
\end{equation}

\noindent
The crucial point is that even for a small mixing parameter $m^2_{12}$
due to the smallness of $(m^2_1-m^2_2)$ the mixing angle $\phi$ is, in general,
not small (at the present state of art, it is impossible to calculate 
the mixing angle $\phi$ reliably). For the most probable case when the
lightest superparticle is the superpartner of the $U(1)$ gauge boson plus some
small mixing with other gaugino and higgsino, the sleptons
$\tilde \mu_R$, $\tilde e_R$ decay mainly into leptons $\mu$ and $e$ plus
$U(1)$ gaugino $\lambda$. The corresponding terms in the Lagrangian 
responsible for slepton decays are

\begin{equation}
L_1~=~\frac{2 g_1}{\sqrt{2}}
(\bar{e}_R \lambda_L \tilde e_R'~+~\bar{\mu}_R \lambda_L \tilde \mu_R'~+~h.c.),
\end{equation}

\noindent
where $g^2_1 \approx 0.13.$ For the case when mixing is absent the decay width
of the right-handed slepton into lepton plus LSP is given by the formulae

\begin{equation}
\Gamma = \frac{g^2_1}{8 \pi} M_{sl} \Delta_f \approx 
5 \cdot 10^{-3} M_{sl}, 
\end{equation}

\begin{equation}
\Delta_f = (1 - \frac{M^2_{LSP}}{M^2_{sl}})^2,
\end{equation}

\noindent
where $M_{sl}$ and $M_{LSP}$ are the masses of the slepton and the lightest
superparticle ($U(1)-$gaugino) respectively. For the case of nonzero
mixing the Lagrangian (10) in terms of the slepton eigenstates reads

\begin{equation}
L_1~=~\frac{2 g_1}{\sqrt{2}}
[\bar{e}_R \lambda_L (\tilde e_R cos(\phi)~-~\tilde{\mu}_R sin(\phi)) +
\bar{\mu}_R \lambda_L (\tilde \mu_R cos(\phi)~-~\tilde{e}_
R sin(\phi)) ~+~h.c.).
\end{equation}

\noindent
Due to nonzero slepton mixing $(sin(\phi) \neq 0)$ we have lepton flavour
number violation in slepton decays, namely:

\begin{equation}
\Gamma(\tilde \mu_R \rightarrow \mu + LSP)~=~\Gamma cos^2(\phi),
\end{equation}

\begin{equation}
\Gamma(\tilde \mu_R \rightarrow e + LSP)~=~\Gamma sin^2(\phi),
\end{equation}

\begin{equation}
\Gamma(\tilde e_R \rightarrow e + LSP)~=~\Gamma cos^2(\phi),
\end{equation}

\begin{equation}
\Gamma(\tilde e_R \rightarrow \mu + LSP)~=~\Gamma sin^2(\phi).
\end{equation}

At LHC the right-handed sleptons are produced mainly through 
the Drell - Yan mechanism
which is flavour blind in such a way that 
even for nonzero slepton mixing, the cross section
$\sigma(p~p~\rightarrow~\tilde \mu^{\pm}_R\tilde e^{\mp}_R+...)$ vanishes
and the single manifestation of the flavour lepton number violation are
sleptons decays with violation of lepton flavour number.

We consider here the most optimistic case of maximal
slepton mixing $(\phi=\frac{\pi}{2})$ and neglect the effects related to 
destructive interference~\cite{14,15,16}. For the case of maximal 
selectron - smuon mixing, the number of signal events coming from slepton
decay is (up to statistical fluctuations)\footnote{As has been mentioned 
above, the contribution of $\tilde \tau_R-$sleptons into the 
$l^+l^-~+~E^{miss}_T$~+~no~jets signature is practically zero and we
neglect it.}

\begin{equation}
N_{sig}(e^+e^-)~=~N_{sig}(\mu^+e^-)~=~N_{sig}(\mu^-e^+)~=~
\frac{1}{4}N^{no~mix}_{sig}(e^+e^-~+~\mu^+\mu^-).
\end{equation}

\noindent
Therefore, for the case of maximal $(\tilde \mu-\tilde e)$ slepton mixing
we expect equal number of signal 
$e^+e^-,~\mu^+\mu^-,~e^+\mu^-,~\mu^+e^-$ events
with $E^{miss}_T$ and with small jet activity. Note that for the
mixing absence case signal events are only $e^+e^-$ and $\mu^+\mu^-$ and,
as a consequence, we have an excess of 
$(e^+e^-+\mu^+\mu^-)$ events over $(e^+\mu^-+\mu^+e^-)$ events due to
nonzero signal events. The number of signal
$(e^+e^-+\mu^+\mu^-+e^+\mu^-+\mu^+e^-)$ events for the case of maximal
mixing coincides (up to statistical fluctuations) with the number of
$(e^+e^-+\mu^+\mu^-)$ signal events without mixing.
Therefore, in our estimates we can use the results of our calculations performed
for the case of zero mixing. We compare the number of background events

\begin{center}
$N^{SM}_B(e^+e^-+\mu^+\mu^-+e^+\mu^-+\mu^+e^-)~=~
2N^{SM}_B(e^+e^-+\mu^+\mu^-)$ 
\end{center}

\noindent
with the number of signal events

\begin{center}
$N_{signal}(e^+e^-+\mu^+\mu^-+e^+\mu^-+\mu^+e^-)~=~
N^{no~mix}_{signal}(e^+e^-+\mu^+\mu^-).$ 
\end{center}

\noindent
The significance is determined by the formulae

\begin{equation}
S = \frac{N^{mix}_{signal}}{\sqrt{N^{mix}_{signal}+N^{mix}_B}}  = 
\frac{N_S}{\sqrt{N_S+2~N^{SM}_B}},
\end{equation}

\noindent
where $N_S$ and $N^{SM}_B$ are the numbers of the signal and the 
background events
for the case of zero mixing. We adopt the standard criterion according to
which the sleptons will be discovered provided the significance is
bigger than $S~\geq~5.$ As it follows from formulae~(19), the maximal
$(\tilde \mu-\tilde e)$ mixing will be discovered provided the significance
of the detection of $(e^+e^-+\mu^+\mu^-)$ events for the case of the mixing
absence is larger than~7. We have found that for the case
of the right-handed sleptons the $(\tilde \mu-\tilde e)$ mixing and,
 hence, flavour
lepton number violation can be detected for the slepton masses up to 250~GeV.
For the case of the left-handed sleptons we can also search for the mixing
effects. In this case, the $(\tilde \mu-\tilde e)$ mixing can also be detected
for the slepton mass up to 250~GeV.

With the maximal stau - smuon mixing the corresponding formulae
are similar to those given above for the selectron - smuon mixing. 
In this case we
expect the number of $e^+e^-$ signal events to be twice greater than the number 
of $\mu^+\mu^-$ signal events and twice smaller than the number of the
$e^+e^-+\mu^+\mu^-$ signal events with no mixing.
Then the significance is

\begin{equation}
S = \frac{N_S(e^+e^-)+N_S(\mu^+\mu^-)}
{\sqrt{N_S(e^+e^-+\mu^+\mu^-)+N_S(e^+e^-)+N_S(\mu^+\mu^-)}}  = 
\frac{{3 \over 4}N_S}{\sqrt{{3 \over 4}N_S+N_B}},
\end{equation}

\noindent
where $N_S$ and $N_B$ are the numbers of signal and background events for the
case of no mixing. Again, if the significance for the case of zero
mixing is larger than~6.6, then the significance~(20) will be larger than~5.

For the case of $(\tilde e - \tilde \tau)$ mixing we do not expect 
$\mu^{\pm}e^{\mp}$ signal events as in the case of the mixing absence.
However, for the case of $(\tilde e - \tilde \tau)$ mixing we expect
the excess of $\mu^+\mu^-$ events over $e^+e^-$ events.

In the Standard Model the difference
$N^{back}(e^+e^-)-N^{back}(\mu^+\mu^-)$ is zero up to statistical
fluctuations. At 1 $\sigma$ level the statistical fluctuation is
$\sqrt{N^{SM}_B}$, where $N^{SM}_B$ is the number of
$e^+e^-+\mu^+\mu^-$ background events, whereas 
$N^{sig}(e^+e^-)-N^{sig}(\mu^+\mu^-)~=~0.25~N^{sig}$ (zero~mixing).
Therefore, it is very difficult to distinguish at the 5 $\sigma$ level between 
the mixing absence case and the $(\tilde \mu-\tilde \tau)$ mixing case.

The case of selectron - stau mixing is similar to that of smuon - stau mixing,
the only difference being the interchange 
$e \rightarrow \mu,~~\mu \rightarrow e.$

For the maximal selectron - smuon - stau mixing, we expect the equal
numbers $e^+e^-,~\mu^+\mu^-,~e^+\mu^-$ and $\mu^+e^-$ signal events.
Therefore, this case is too similar to that of the maximal 
$(\tilde \mu-\tilde e)$ mixing. The sole difference is the number 
of signal events 

\begin{center}
$N_S(e^+e^-+\mu^+\mu^-+e^+\mu^-+e^-\mu^+)~=
~\frac{2}{3}N^{zero~mixing}_S(e^+e^-+\mu^+\mu^-).$
\end{center}

\noindent
So, the significance is
$S~=~\frac{{2 \over 3}N_S}{\sqrt{2N_B+N_S}},$ where $N_S$ and $N_B$
is the number of signal events for the case of zero mixing.
The significance $S$ for the maximal
$(\tilde \mu-\tilde e-\tilde \tau)$ mixing exceeds 5 provided
the corresponding significance for the case of zero mixing 
is larger than~10. So, at CMS it would be extremely difficult to detect
the $(\tilde \mu-\tilde e-\tilde \tau)$ mixing.

\section{Conclusion}

We have studied separetely
the possibility to detect the right-handed sleptons, the left-handed sleptons 
and right- plus left-handed sleptons at CMS in a model independent way.
     
For the right-handed sleptons the number of signal events passing through cuts
depends on the mass of the slepton and the mass of the LSP. We have found that 
for luminosity 
$L~=~10^5pb^{-1}$ it would be possible to discover the right-handed
sleptons for a mass up to 325~GeV using the standard significance criterion
$S~=~\frac{N_S}{\sqrt{N_S+N_B}}~\geq~5.$ However, the SM background has 
an equal number of $(e^+e^-+\mu^+\mu^-)$ and $(e^+\mu^-+\mu^+e^-)$ events 
and the signal contributes only to $(e^+e^-+\mu^+\mu^-)$ events, 
so we can estimate the difference
$\Delta N~=~N(e^+e^-+\mu^+\mu^-)~-~N(e^+\mu^-+\mu^+e^-).$ Only signal
events contribute to $\Delta N.$ In the Standard Model $\Delta N$ is 
equal to zero up to statistical fluctuations.
Nonzero $\Delta N$ is an independent and very important check
for the slepton discovery at LHC. For the case when only the left-handed
sleptons contribute to signal events, we have found that it is possible
to discover the left-handed sleptons with a mass up to 350~GeV.

For the right-handed sleptons we have found that the number of signal events
decreases with the increase of the LSP mass and, typically, it is possible 
to detect the right-handed sleptons provided the LSP mass 
$m_{LSP}~\leq~0.4~m_{\tilde l_R}.$

For the left-handed sleptons we have found that the number of the signal
events is maximal for $m_{LSP}~=~(0.4-0.6)~m_{\tilde l_L}.$
For the LSP masses in this interval, the CMS left-handed slepton discovery
potential is the maximal one.  Note that these results are in agreement
with the similar observations of ref.~\cite{3}. 

We have also studied 
the case of flavour lepton number violation in slepton decays. For
the case of maximal $(\tilde \mu_R-\tilde e_R)$ mixing we have found 
that the signature qualitatively differs from the case of zero mixing,
namely, in this case we do not have an excess of 
$\Delta N~=~N(e^+e^-+\mu^+\mu^-)~-~N(e^+\mu^-+\mu^+e^-)$ events unlike
the case of zero mixing where $\Delta N~>~0.$ So, it is possible 
to distinguish zero mixing and maximal mixing. We have found that it is 
possible to detect the maximal $(\tilde \mu_R-\tilde e_R)$ mixing for the 
right-handed sleptons with a mass up to 250~GeV. We also considered 
the cases of $(\tilde \mu-\tilde \tau)$ and 
$(\tilde \mu-\tilde e-\tilde \tau)$ mixings. However, for such mixing at
$L~=~10^5pb^{-1}$ it is not so easy to distinguish the  mixings from
the case of the mixing absence. Our conclusion about the possibility
to detect sleptons with a mass up to 300~GeV is in
qualitative agreement with the similar results of ref.~\cite{3}. However,
in our paper we have studied a more general situation (we have not assumed
MSUGRA model, which as it has been explained in the Introduction is, 
at best, only a rough description of  the sparticle spectrum). In fact,
the number of signal events passing the cuts depends mainly on the masses
of the right- and the left-handed sleptons and on the LSP mass. So, those 
3 parameters determine the possibility to detect sleptons at CMS.
As a rule, we neglecte cascade neutralino or chargino decays 
resulting in the dilepton signature. However, we have checked that  
(especially for cuts 3-7) their contribution is generally not very large
and, moreover, an account of such contribution increases the significance. 
The reason why we have neglected gaugino decays is that, in general, the masses
of $\tilde \chi^0_2,~\tilde \chi^{\pm}_1$ are model dependent (they  are determined from
standard but an "ad hoc" assumption that at GUT scale all gaugino masses 
coincide).

\begin{center}
 {\large \bf Acknowledgments}
\end{center}

\par
We are  indebted to the participants of Daniel Denegri seminar on physics 
simulations for useful discussions. We would like to thank Luly Rurua
for providing us her code of the events selections. We are indepted to
Michael Dittmar for a critical reading of the manuscript and for valuable
comments.
\par
The research described in this publication was made possible in part by 
Award No RP1-187 of the U.S. Civilian Research and Development Foundation for 
the Independent States of the Former Soviet Union(CRDF).

\bigskip

\newpage

\begin{table}[t]
\small
    \caption{The number of events $N_{ev}$ and significance 
$S$ for the case of right-handed sleptons, 
$m_{\tilde l_R}$~=~96~GeV,~L~=~$10^{4}$~pb$^{-1}$
and for different LSP masses $m_{\tilde \chi^0_1}$. Columns for sets 3-7 
contain insignificant information. }
    \label{tab.4}
\begin{center}
\begin{tabular}{|l|l|l|l|}
\hline
 $m_{\tilde \chi^0_1}$ &   & cut 1 & cut 2 \\
\hline
24~GeV  & $N_{ev}$  & 243   &   463 \\
        &  $S$      & 6.9   &   8.6 \\
\hline
38~GeV  & $N_{ev}$ & 89    &   180 \\
       & $S$      & 2.7   &   3.5 \\
\hline
53~GeV & $N_{ev}$ & 34    &    34 \\
       & $S$      & 2.7   &   3.5 \\
\hline
 & $N^{SM}_B$~\cite{3} & 992 & 2421 \\
\hline
\end{tabular}
\end{center}
\end{table}

\begin{table}[b]
\small
 \caption{The number of events $N_{ev}$ and significance 
$S$ for the case of right-handed sleptons, 
$m_{\tilde l_R}$~=~100~GeV,~L~=~$10^{4}$~pb$^{-1}$
and for different LSP masses $m_{\tilde \chi^0_1}$.}
    \label{tab:Tab.5}
\begin{center}
\begin{tabular}{|l|l|l|l| }
\hline
$m_{\tilde \chi^0_1}$~&    & cut 1 & cut 2 \\
\hline
24~GeV & $N_{ev}$   & 195.  &   366.\\
       & $S$        & 5.7   &   6.9 \\
\hline
 38~GeV & $N_{ev}$  &     171. &    316. \\
        &  $S$      &     5.0  &     6.0 \\
\hline
53~GeV  &  $N_{ev}$ &     120. &    219. \\
        &  $S$      &     3.6  &    4.3 \\
\hline
69~GeV &  $N_{ev}$  &     48.  &    79. \\
       &   $S$      &     1.5  &    1.6 \\
\hline
\end{tabular}
\end{center}
\end{table}

\begin{table}[t]
\small
    \caption{The number of events $N_{ev}$ and significance 
$S$ and for the case of right-handed sleptons, 
$m_{\tilde l_R}~=~125$~GeV, $L~=~10^{5}$~pb$^{-1}$
and for different LSP masses $m_{\tilde \chi^0_1}$.}
    \label{tab:Tab.6}
\begin{center}
\begin{tabular}{|l|l|l|l|l|l|l|l|l| }
\hline
$ m_{\tilde \chi^0_1}$ & & cut 1 & cut 2 & cut 3 & cut 4 & cut 5 & cut 6 & cut 7 \\
\hline
26~GeV& $N_{ev}$ & 1086.& 2091.& 189.& 79. & 12. & 27. & 12. \\
&$ S $           & 10.4 & 12.9 & 9.9 & 5.8 & 1.6 & 3.0 & 1.7 \\
\hline
54~GeV&$N_{ev}$  & 806. & 1632.& 50. & 19. &  0. & 6.  &  1. \\
&$ S $           &  7.8 & 10.2 & 3.4 & 1.7 & 0.0 & 0.8 & 0.2 \\
\hline
85~GeV&$N_{ev}$  & 446. & 845. &  0. &  0. &  0. &  0. &  0. \\
&$ S $           &  4.4 & 5.3  & 0.0 & 0.0 & 0.0 & 0.0 &  0.0 \\
\hline
 & $N^{SM}_B$~\cite{3}&9920 &24210 & 172 & 105 &  45 &  53 &38 \\
\hline
\end{tabular}
\end{center}
  \end{table}

\begin{table}[b]
\small
    \caption{The number of events $N_{ev}$ and significance 
$S$ for the case of right-handed sleptons, 
$m_{\tilde l_R}$~=~150~GeV, $L~=~10^{5}$~pb$^{-1}$
and for different LSP masses $m_{\tilde \chi^0_1}$.}
    \label{tab:Tab.7}
\begin{center}
\begin{tabular}{|l|l|l|l|l|l|l|l|l| }
\hline
$ m_{\tilde \chi^0_1}$ & & cut 1 & cut 2 & cut 3 & cut 4 & cut 5 & cut 6 & cut 7 \\
\hline
24~GeV& $N_{ev}$ & 626.&1209.& 185.& 127.&  38.&  60.&  38. \\
&$ S $                 & 6.1 & 7.6 & 9.8 & 8.3 & 4.2 & 5.6 &  4.4 \\
\hline
53~GeV& $N_{ev}$ & 595.&1183.& 149.&  83.&  15.&  23.&  15. \\
&$ S $                 & 5.8 & 7.4 & 8.3 & 6.1 & 1.9 & 2.6 &  2.1 \\
\hline
69~GeV& $N_{ev}$ & 472.& 922.& 23. & 5.  &  0. & 1.  &  0.  \\
&$ S $                 &  4.6& 5.8 & 1.6 & 0.5 & 0.0 & 0.1 & 0.0  \\
\hline
\end{tabular}
\end{center}
  \end{table}

\begin{table}[t]
\small
    \caption{The number of events $N_{ev}$ and significance 
$S$ for the case of right-handed sleptons, 
$m_{\tilde l_R}$~=~175~GeV, $L~=~10^{5}$~pb$^{-1}$
and for different LSP masses $m_{\tilde \chi^0_1}$.}
    \label{tab:Tab.8}
\begin{center}
\begin{tabular}{|l|l|l|l|l|l|l|l|l| }
\hline
$ m_{\tilde \chi^0_1}$ & & cut 1 & cut 2 & cut 3 & cut 4 & cut 5 & cut 6 & cut 7 \\
\hline
26~GeV& $N_{ev}$  & 345.& 679.& 155.& 130.&  73.&  92.&  72. \\
&$ S $                 & 3.4 &  4.3& 8.6 & 8.5 & 6.7 & 7.6 & 6.9 \\
\hline
54~GeV& $N_{ev}$  & 289.& 648.& 137.& 108.& 49. &  61.&  49. \\
&$ S $                 & 2.9 & 4.1 & 7.8 & 7.4 & 5.1 & 5.7 & 5.3 \\
\hline
85~GeV& $N_{ev}$  & 317.& 647.&  83.&  47.&   7.&  17.&  7. \\
&$ S $                 & 3.1 &  4.1& 5.2 & 3.8 & 1.0 & 2.0 & 1.0 \\
\hline
\end{tabular}
\end{center}
  \end{table}

\begin{table}[b]
\small
    \caption{The number of events $N_{ev}$  and significance 
$S$ for the case of right-handed sleptons, 
$m_{\tilde l_R}$~=~200~GeV, $L~=~10^{5}$~pb$^{-1}$
and for different LSP masses $m_{\tilde \chi^0_1}$.}
    \label{tab:Tab.9}
\begin{center}
\begin{tabular}{|l|l|l|l|l|l|l|l|l| }
\hline
 $m_{\tilde \chi^0_1}$ & & cut 1 & cut 2 & cut 3 & cut 4 & cut 5 & cut 6 & cut 7 \\
\hline
24~GeV& $N_{ev}$  & 229.& 506.& 170.& 152.& 110.& 128.& 110. \\
&$ S $                 & 2.3 & 3.2 & 9.2 & 9.5 & 8.8 & 9.5 & 9.0 \\
\hline
53~GeV& $N_{ev}$  & 248.& 476.& 117.& 106.& 75. & 96. & 75. \\
&$ S $                 & 2.5 & 3.0 & 6.9 & 7.3 & 6.8 & 7.9 & 7.1 \\
\hline
85~GeV& $N_{ev}$  & 231.& 447.&  90.&  73.&  40.&  55.&  40. \\
&$ S $                 & 2.3 & 2.8 & 5.6 & 5.5 & 4.3 & 5.3 & 4.5 \\
\hline
119~GeV& $N_{ev}$  &  81.& 175.&  1. &  0. &  0. &  0. &   0. \\
&$ S $                 & 0.8 & 1.1 & 0.1 & 0.0 & 0.0 & 0.0 &  0.0 \\
\hline
\end{tabular}
\end{center}
  \end{table}

\begin{table}[t]
\small
    \caption{The number of events $N_{ev}$ and significance 
$S$ for the case of right-handed sleptons, 
$m_{\tilde l_R}~=~225~GeV,$ $L~=~10^{5}pb^{-1}$
and for different LSP masses $m_{\tilde \chi^0_1}$.}
    \label{tab:Tab.10}
\begin{center}
\begin{tabular}{|l|l|l|l|l|l|l|l|l| }
\hline
$ m_{\tilde \chi^0_1}$&  & cut 1 & cut 2 & cut 3 & cut 4 & cut 5 & cut 6 & cut 7 \\
\hline
26~GeV&$N_{ev}$ & 150.& 268.&  81.&  76.&  69.&  87.&  69. \\
&$ S $                 & 1.5 & 1.7 & 5.1 & 5.6 & 6.5 & 7.4 & 6.7 \\
\hline
54~GeV&$N_{ev}$ & 140.& 264.&  84.&  82.&  63.&  75.&  62. \\
&$ S $                 & 1.4 & 1.7 & 5.3 & 6.0 & 6.1 & 6.6 & 6.2 \\
\hline
85~GeV&$N_{ev}$  & 156.& 306.&  85.&  78.&  54.&  66.&  53. \\
&$ S $                 & 1.6 & 2.0 & 5.3 & 5.8 & 5.4 & 6.1 & 5.6 \\
\hline
119~GeV&$N_{ev}$ & 135.& 277.&  68.&  55.&  28.&  33.&  28. \\
&$ S $                 & 1.3 & 1.8 & 4.4 & 4.3 & 3.3 & 3.6 & 3.4 \\
\hline
\end{tabular}
\end{center}
  \end{table}

\begin{table}[b]
\small
    \caption{The number of events $N_{ev}$  and significance 
$S$ for the case of right-handed sleptons, 
$m_{\tilde l_R}$~=~250~GeV, $L~=~10^{5}$~pb$^{-1}$
and for different LSP masses $m_{\tilde \chi^0_1}$.}
    \label{tab:Tab.11}
\begin{center}
\begin{tabular}{|l|l|l|l|l|l|l|l|l| }
\hline
$ m_{\tilde \chi^0_1}$&  & cut 1 & cut 2 & cut 3 & cut 4 & cut 5 & cut 6 & cut 7 \\
\hline
24~GeV& $N_{ev}$  & 128.& 237.&  84.&  78.&  81.&  92.&  81. \\
&$ S $                 & 1.3 & 1.5 & 5.3 & 5.8 & 7.2 & 7.6 & 7.4 \\
\hline
53~GeV&  $N_{ev}$ & 125.& 232.&  82.&  76.&  78.&  91.&  78. \\
&$ S $                 & 1.2 & 1.5 & 5.1 & 5.6 & 7.0 & 7.6 &  7.2 \\
\hline
85~GeV& $N_{ev}$  & 117.& 220.&  78.&  73.&  68.&  80.&  68. \\
&$ S $                 & 1.2 & 1.4 & 4.9 & 5.5 & 6.4 & 6.9 & 6.6 \\
\hline
119~GeV& $N_{ev}$  & 116.& 217.&  66.&  61.&  49.&  56.&  49. \\
&$ S $                 & 1.2 & 1.4 & 4.3 & 4.7 & 5.1 & 5.4 &  5.3 \\
\hline
157~GeV& $N_{ev}$  & 94. & 187.& 43. &  31.&  15.&  18.&  15. \\
&$ S$                  & 0.9 & 1.2 & 2.9 & 2.7 & 1.9 & 2.1 & 2.1 \\
\hline
\end{tabular}
\end{center}
  \end{table}

\begin{table}[t]
\small
    \caption{The number of events $N_{ev}$ and significance 
$S$ for the case of right-handed sleptons, 
$m_{\tilde l_R}$~=~275~GeV, $L~=~10^{5}$~pb$^{-1}$
and for different LSP masses $m_{\tilde \chi^0_1}$.}
    \label{tab:Tab.12}
\begin{center}
\begin{tabular}{|l|l|l|l|l|l|l|l|l| }
\hline
$ m_{\tilde \chi^0_1}$&  & cut 1 & cut 2 & cut 3 & cut 4 & cut 5 & cut 6 & cut 7 \\
\hline
25~GeV&$N_{ev}$  &  73.& 130.&  45.&  44.&  40.&  49.&  40. \\
&$ S $                 & 0.7 & 0.8 & 3.1 & 3.6 & 4.3 & 4.9 &  4.5 \\
\hline
54~GeV&$N_{ev}$  &  68.& 139.&  53.&  52.&  49.&  59.&  49. \\
&$ S $                 & 0.7 & 0.9 & 3.5 & 4.2 & 5.1 & 5.6 & 5.3 \\
\hline
85~GeV&$N_{ev}$  &  60.& 115.&  35.&  32.&  31.&  34.&  31. \\
&$ S$                  & 0.6 & 0.7 & 2.4 & 2.7 & 3.6 & 3.6 & 3.7 \\
\hline
119~GeV&$N_{ev}$  &  91.& 169.&  56.& 54. & 47. & 57. &  46. \\
&$ S $                 & 0.9 & 1.1 & 3.7 & 4.3 & 4.9 & 5.4 & 5.0 \\
\hline
\end{tabular}
\end{center}
  \end{table}

\begin{table}[b]
\small
    \caption{The number of events $N_{ev}$  and significance 
$S$ for the case of right-handed sleptons, 
$m_{\tilde l_R}$~=~300~GeV, $L~=~10^{5}$~pb$^{-1}$
and for different LSP masses $m_{\tilde \chi^0_1}$.}
    \label{tab:Tab.13}
\begin{center}
\begin{tabular}{|l|l|l|l|l|l|l|l|l| }
\hline
$ m_{\tilde \chi^0_1}$&  & cut 1 & cut 2 & cut 3 & cut 4 & cut 5 & cut 6 & cut 7 \\
\hline
24~GeV&  $N_{ev}$&  64.& 134.&  59.&  59.&  56.&  65.&  56. \\
&$ S $                 & 0.6 & 0.9 & 3.9 & 4.6 & 5.6 & 6.0 & 5.8 \\
\hline
52~GeV& $N_{ev}$ &  59.& 117.&  50.&  46.&  45.&  53.&  45. \\
&$ S$                  & 0.6 & 0.8 & 3.4 & 3.7 & 4.7 & 5.1 & 4.9 \\
\hline
85~GeV& $N_{ev}$ &  55.& 118.&  49.&  46.&  45.&  54.&  45. \\
&$ S $                 & 0.6 & 0.8 & 3.3 & 3.7 & 4.7 &  5.2&  4.9 \\
\hline
119~GeV& $N_{ev}$ &  56.& 114.&  46.&  44.&  41.&  48.&  41. \\
&$ S $                 & 0.6 & 0.7 & 3.1 & 3.6 & 4.4 & 4.8 & 4.6 \\
\hline
157~GeV& $N_{ev}$ &  54.& 112.&  38.&  36.&  33.&  40.&  33. \\
&$ S $                 & 0.5 & 0.7 & 2.6 & 3.0 & 3.7 & 4.1 & 3.9 \\
\hline
196~GeV& $N_{ev}$ &  52.& 105.&  29.&  25.&  13.&  16.&  13. \\
&$ S  $                & 0.5 & 0.7 & 2.0 & 2.2 & 1.7 & 1.9 &  1.8 \\
\hline
\end{tabular}
\end{center}
  \end{table}

\begin{table}[t]
\small
    \caption{The number of events $N_{ev}$  and significance 
$S$ for the case of right-handed sleptons, 
$m_{\tilde l_R}$~=~325~GeV, $L~=~10^{5}$~pb$^{-1}$
and for different LSP masses $m_{\tilde \chi^0_1}$.}
    \label{tab:Tab.14}
\begin{center}
\begin{tabular}{|l|l|l|l|l|l|l|l|l| }
\hline
$ m_{\tilde \chi^0_1}$&  & cut 1 & cut 2 & cut 3 & cut 4 & cut 5 & cut 6 & cut 7 \\
\hline
53~GeV& $N_{ev}$  &  24.&  72.&  35.&  35.&  36.&  45.&  35. \\
&$ S  $                & 0.2 & 0.5 & 2.4 & 3.0 & 4.0 & 4.5 &  4.1 \\
\hline
85~GeV& $N_{ev}$  &  44.&  89.&  39.&  38.&  40.&  51.&  40. \\
&$ S  $                & 0.4 & 0.6 & 2.7 & 3.2 & 4.3 & 5.0 & 4.5  \\
\hline
119~GeV& $N_{ev}$  &  32.&  74.&  37.&  36.&  33.&  43.&  33. \\
&$ S  $                & 0.3 & 0.5 & 2.6 & 3.0 & 3.7 & 4.4 & 3.9  \\
\hline
157~GeV& $N_{ev}$  &  34.&  76.&  31.&  29.&  28.&  37.&   28. \\
&$ S  $                & 0.3 & 0.5 & 2.2 & 2.5 & 3.3 & 3.9 &  3.4  \\
\hline
196~GeV& $N_{ev}$  &  32.&  73.&  28.&  26.&  19.&  26.&   19. \\
&$ S $                 & 0.3 & 0.5 & 2.0 & 2.3 & 2.4 & 2.9 & 2.5 \\
\hline
233~GeV& $N_{ev}$  &  30.&  62.&  17.&  13.&   4.&   6.&   4. \\
&$ S $                 & 0.3 & 0.4 & 1.2 & 1.2 & 0.6 & 0.8 &  0.6 \\
\hline
\end{tabular}
\end{center}
  \end{table}

\begin{table}[t]
\small
    \caption{The number of events $N_{ev}$  and significance 
$S$ for the case of right-handed sleptons, 
$m_{\tilde l_R}$~=~350~GeV, $L~=~10^{5}$~pb$^{-1}$
and for different LSP masses $m_{\tilde \chi^0_1}$.}
    \label{tab:Tab.15}
\begin{center}
\begin{tabular}{|l|l|l|l|l|l|l|l|l| }
\hline
$ m_{\tilde \chi^0_1}$&  & cut 1 & cut 2 & cut 3 & cut 4 & cut 5 & cut 6 & cut 7 \\
\hline
53~GeV& $N_{ev}$ &  33.&  68.&  33.&  33.&  33.&  40.&  33. \\
&$ S  $                & 0.3 & 0.4 & 2.3 & 2.8 & 3.7 & 4.1 &  3.9 \\
\hline
119~GeV& $N_{ev}$ & 32. &  69.& 35. & 34. & 35. & 36. &  35. \\
&$ S $                 & 0.3 & 0.4 & 2.4 & 2.9 & 3.9 & 3.8 & 4.1 \\
\hline
196~GeV& $N_{ev}$ &  30.&  65.&  31.&  31.&  27.&  27.&  27. \\
&$ S$                  & 0.3 & 0.4 & 2.2 & 2.7 & 3.2 & 3.0 & 3.3 \\
\hline
233~GeV& $N_{ev}$ &  27.&  61.&  25.&  23.&  18.&  18.&  18. \\
&$ S $                 & 0.3 & 0.4 & 1.8 & 2.0 & 2.3 & 2.1 &  2.4 \\
\hline
270~GeV& $N_{ev}$ &  23.&  56.&  12.&   7.&   3.&   3.&   3. \\
&$ S  $                & 0.2 & 0.4 & 0.9 & 0.7 & 0.4 & 0.4 & 0.5 \\
\hline
\end{tabular}
\end{center}
  \end{table}

\begin{table}[b]
\small
\caption{The number of events $N_{ev}$  and significance 
$S$ for the case of left-handed sleptons, 
$m_{\tilde l_L}~=~100$~GeV,~~L~=~10$^{4}$~pb$^{-1}$
and for different LSP masses $m_{\tilde \chi^0_1}$. Columns for sets 3-7 
contain insignificant information.}
\label{tab:Tab.16}
\begin{center}
\begin{tabular}{|l|l|l|l| }
\hline
$ m_{\tilde \chi^0_1}$ & & cut 1 & cut 2 \\
\hline
24~GeV& $N_{ev}$  & 132.& 546. \\
&$ S  $                & 3.9 & 10.0 \\
\hline
38~GeV& $N_{ev}$  & 122.& 372. \\
&$ S  $                & 3.7 & 7.0 \\
\hline
53~GeV& $N_{ev}$  & 421. & 602. \\
&$ S  $                & 11.2 & 10.9 \\
\hline
69~GeV& $N_{ev}$  & 209. & 291. \\
&$ S  $                & 6.0  & 5.6 \\
\hline
 & $N^{SM}_B$~\cite{3}  & 992  &2421 \\
\hline
\end{tabular}
\end{center}
\end{table}

\begin{table}[t]
\small
\caption{The number of events $N_{ev}$  and significance 
$S$ for the case of left-handed sleptons, 
$m_{\tilde l_L}$~=~150~GeV,~~L~=~$10^{5}$~pb$^{-1}$
and for different LSP masses $m_{\tilde \chi^0_1}$.}
    \label{tab:Tab.17} 
\begin{center}
\begin{tabular}{|l|l|l|l|l|l|l|l|l|}
\hline
$ m_{\tilde \chi^0_1}$ &  & cut 1 & cut 2 & cut 3 & cut 4 & cut 5 & cut 6 & cut 7 \\
\hline
24~GeV&$N_{ev}$ & 682.&1815.& 302.& 131.&  32.&  41.&  32. \\
&$ S $                 &  6.6& 11.3& 13.9& 8.5 & 3.6 & 4.2 & 3.8 \\
\hline
53~GeV&$N_{ev}$ & 663.&1762.& 143.& 111.& 113.& 120.& 113. \\
&$ S $                 & 6.4& 10.9& 8.1 & 7.6 & 9.0 & 9.1 & 9.2 \\
\hline
85~GeV&$N_{ev}$ & 951.&2029.&  72.&  11.&   2.&  23.&  2. \\
&$ S $                 &  9.1& 12.5& 4.6 & 1.0 & 0.3 & 2.6 & 0.3 \\
\hline
119~GeV&$N_{ev}$ & 273.& 485.&   2.&   1.&   1.&   2.&  1. \\
&$ S $                 & 2.7 & 3.1 & 0.2 & 0.1 & 0.1 &  0.3&  0.2 \\
\hline
 & $N^{SM}_B$~\cite{3}  &9920 &24210 & 172 & 105 &  45 &  53 &38 \\
\hline
\end{tabular}
\end{center}
\end{table}

\begin{table}[b]
\small
    \caption{The number of events  $N_{ev}$ and significance 
$S$ for the case of left-handed sleptons, 
$m_{\tilde l_L}~=~200$~GeV,~~L~=~10$^{5}$~pb$^{-1}$
and for different LSP masses $m_{\tilde \chi^0_1}$.}
\label{tab:Tab.18}
\begin{center}
\begin{tabular}{|l|l|l|l|l|l|l|l|l|}
\hline
$ m_{\tilde \chi^0_1}$&  & cut 1 & cut 2 & cut 3 & cut 4 & cut 5 & cut 6 & cut 7 \\
\hline
53~GeV&$N_{ev}$ & 202.& 618.& 117.&  88.&  49.&  51.&  49. \\
&$ S $                 & 2.0 & 3.9 & 6.9 & 6.3 & 5.1 & 5.0 & 5.3 \\
\hline
85~GeV&$N_{ev}$ & 252.& 641.& 108.&  86.&  43.&  54.&  43. \\
&$ S $                 & 2.5 &  4.1& 6.5 & 6.2 & 4.6 &  5.2&  4.8 \\
\hline
119~GeV&$N_{ev}$ & 468.& 920.&  91.&  42.&  11.&  14.&  11. \\
&$ S $                 &  4.6&  5.8& 5.6 & 3.5 & 1.5 & 1.7 & 1.6  \\
\hline
\end{tabular}
\end{center}
\end{table}

\end{document}